# Fabrication of high quality GaN nanopillar arrays by dry and wet chemical etching

Running title: Fabrication of high quality GaN nanopillar

Running Authors: Paramanik et al.

Dipak Paramanik,[1,a)] Abhishek Motayed,[1,2,] Matthew King,[3] Jong-Yoon Ha,[1,2] Sergi Kryluk,[1,2] Albert V. Davydov[1] Alec Talin,[1]

[1]National Institute of Standards and Technology, Gaithersburg, Maryland 20899, USA

[2]Institute for Research in Electronics and Applied Physics, University of Maryland, College Park, Maryland 20742, USA

[3]Northrop Grumman ES, Linthicum, Maryland 21090, USA

Corresponding author, Electronic mail: dipakiop@gmail.com,

We study strain relaxation and surface damage of GaN nanopillar arrays fabricated using inductively coupled plasma (ICP) etching and post etch wet chemical treatment. We controlled the shape and surface damage of such nanopillar structures through selection of etching parameters. We compared different substrate temperatures and different chlorine-based etch chemistries to fabricate high quality GaN nanopillars. Room temperature photoluminescence and Raman scattering measurements were carried to study the presence of surface defect and strain relaxation on these nanostructures, respectively. We found that wet KOH etching can remove the side wall damages caused by dry plasma etching, leading to better quality of GaN nanopillars arrays. The Si material underneath the GaN pillars was removed by KOH wet etching, leaving behind a fine Si pillar to support the GaN structure. Substantial strain relaxations were observed in these structures from room temperature Raman spectroscopy measurements. Room temperature Photoluminescence spectroscopy shows the presence of whispering gallery modes from these the nano disks structures.

## I. INTRODUCTION

Top down fabrication method would be suitable for producing semiconductor nanopillar arrays for device application only when four basic requirements are satisfied: (1) control over shape and morphology, (2) defect free smooth surface of side wall and base plane, (3) ability to tailor profile for specific application, and (4) large-area uniformity and scalability. Development of nanopillar arrays with high aspect ratios (10 and higher) where the diameter are in the range of 150-250 nm require careful design of the etch process and selection of the mask material. It is also essential to analyze the sub-surface defect as well strain/stress of the fabricated nanopillars arrays. Optimum dry etching condition and post wet chemical etching can minimize the surface defect and relax the strain of the nano pillars. Smooth and defect free sidewall is important when the etched pillars are used as a core for the growth of n-core/p-shell type structure for vertical p-n junction devices. In the case of vertical transistor and light emitting diode , a straight profile with 90° angle is highly desirable, whereas for field emitter a slopped profile with sharp tip is needed. Direct band-gap gallium nitride (GaN) nanopillar arrays have attracted tremendous research interest for potential device applications.[1-4] Using top-down approach, uniform GaN nanopillar arrays have been fabricated through a combination of lithography and large area etching using inductively coupled plasma (ICP) tools.[5-9] In our recent studies it has been demonstrated that lowering the ion energy by lowering the RF power or increasing the chemical activity by changing the gas chemistry or increasing the ICP power in the plasma to minimize the etch damage often results in slower etch rates and less anisotropic profiles which significantly limits figure of merit of the fabricated nanostructures.[10] Cheung at el,[11] found that higher concentration of donor-related defects are introduced on the top 100 nm GaN surface after Ar compared to $SF_6$ plasma treatment. So far, only few data have been reported concerning plasma-induced damage in GaN nanopillars.[12-15]

In this article, we report the changes in surface damage of GaN nanopillar fabricated using different etch conditions and post wet chemical etching. The surface damage and strain relaxation of these nano pillar have been studied using room temperature photoluminescence (PL) and Raman scattering techniques. KOH wet chemical have been used to remove the plasma-damaged material and to improve the optical quality of the fabricated nanopillars. KOH etching has also been utilized to remove the Si under the GaN pillar and to fabricate nonodisk structures.

## II. EXPERIMENTAL

The GaN samples used for this study were grown on n-type Si (111) substrates with low sheet resistance of 0.005 Ω/square. A commercial metal organic chemical vapor deposition system was used to grow the GaN epitaxial layers. The thickness of the GaN epilayer was 0.8 $\mu$m. Intermediate $Al_{1-x}Ga_xN$ buffer layers with varying x and thickness of 150 nm were utilized.

After the growth, the GaN/AlGaN/Si wafers were patterned using deep UV lithography. The etch mask used for this study was Ti/Ni (50 nm/120 nm) deposited by e-beam evaporation and patterned by liftoff. The diameters of the etch mask was ranging from 10 $\mu$m to 250 nm and pitches ranging from 20 $\mu$m to 100 nm. For the etching experiments, the wafers were diced into 20 mm × 20 mm pieces and mounted on a 4 inch Si carrier wafer with 50 nm atomic layer deposited $Al_2O_3$. Thermal contact to the carrier wafer was made with suitable thermal grease. The samples were etched in an Oxford PlasmaLab 100 ICP system with a Oxford remote ICP380 source.[16] The helium pressure for wafer backside cooling was 1.33 kPa (10 Torr) and flow was about 1.5 sccm. For every etch process the dc self-bias was recorded. Before every etch experiment, the ICP chamber was cleaned and conditioned. After the ICP etching, the samples were put in $HF:HNO_3:H_2O$ (1:1:10) solution for 2 min to remove the etch mask and clean both etch debris and redeposit. The samples were then observed with a 70° tilted sample-holder in a field-emission scanning electron microscope (FESEM). In order to compare results for the different etch conditions, we computed the height and sidewall angle, $\theta$, for every nanopillar sample. Room-temperature photoluminescence (PL), and Raman scattering spectroscopy was conducted on the samples using 325 nm laser and 650 nm laser, respectively on an integrated system. The damaged materials from the side wall of the nanopillars were removed by wet chemical etching using 10 % KOH solution in water as well as in ethylene glycol (EG). The temperature KOH solution was always kept fixed at 40 °C during the wet etching.

## III. RESULTS AND DISCUSSION

Fig.1 (a to c) presents the FESEM images of 250 nm diameter starting circle pattern etched using Cl2/N2/Ar (25/5/2 sccm) chemistry at three different substrate temperature ranging from low at -120 °C, mid 40 °C and high 350 °C. The other etch parameter ICP power 800 W, RF power 300 W, pressure 5 mT and etching time 5 min were kept constant. The four numbers on each figure are as follows: 1) diameter at the top of the pillar in nm, 2) diameter at the bottom in nm, 3) on the left side total height of the pillar in nm, and 4) angle $\theta$ of the sidewall measured in degrees at the top left corner of the imageAlthough the sample were mounted on the Si career wafer it is expected that the sample will reach thermal equilibrium with the cathode due to long wait time between loading and actual etch. Fig1 shows the significant change in the side wall angles of the fabricated nano structure at with temperaures. Fig 1a. show the tapered cone shape nanopillar with top diameter 30 nm and bottom diameter 680 nm with sloped sidewall angle 75° . This type of nano cones were formed for low temperature etching at -40 °C. As the temperature increases the vertical etching rate increases slightly as it is clear from the height of the nanopillars labeled in fig 1. However the significant lateral etching of the mask at low temperature may be the cause for the formation of sharp tip in fig 1a. The lateral mask erosion at lower temperature is surprising. Also it is interesting that to note that the sidewall angle increases with increasing temperature. Slopped sidewalls at lower temperatures indicate a sputter-dominated regime with limited volatility of etch products. At lower cathode temperatures, the etch product ($GaCl_3$) might be solid as the melting point is 77.9 °C at atmospheric pressure. At higher temperature the rate of chemical etching increases which enhance the lateral etching of the sidewalls. Thus a vertical nanopillars with higher sidewall angles were formed for high temperature etching as shown in fig.1c. In our earlier studies,[10] it was reported that a significant tapering of the etch profile can be obtained by changing the etch chemistry from pure $Cl_2$ to pure Ar. Since Ar can cause only the physical etching, the chemical etching is necessary to do the lateral etch along the height of the pillar. Since the chemical activity was more at higher temperature and hence the lateral etching was more, as a result nearly vertical (θ = 88°) nanopillars were formed at temperature 350 °C (fig. 1 c, d ).

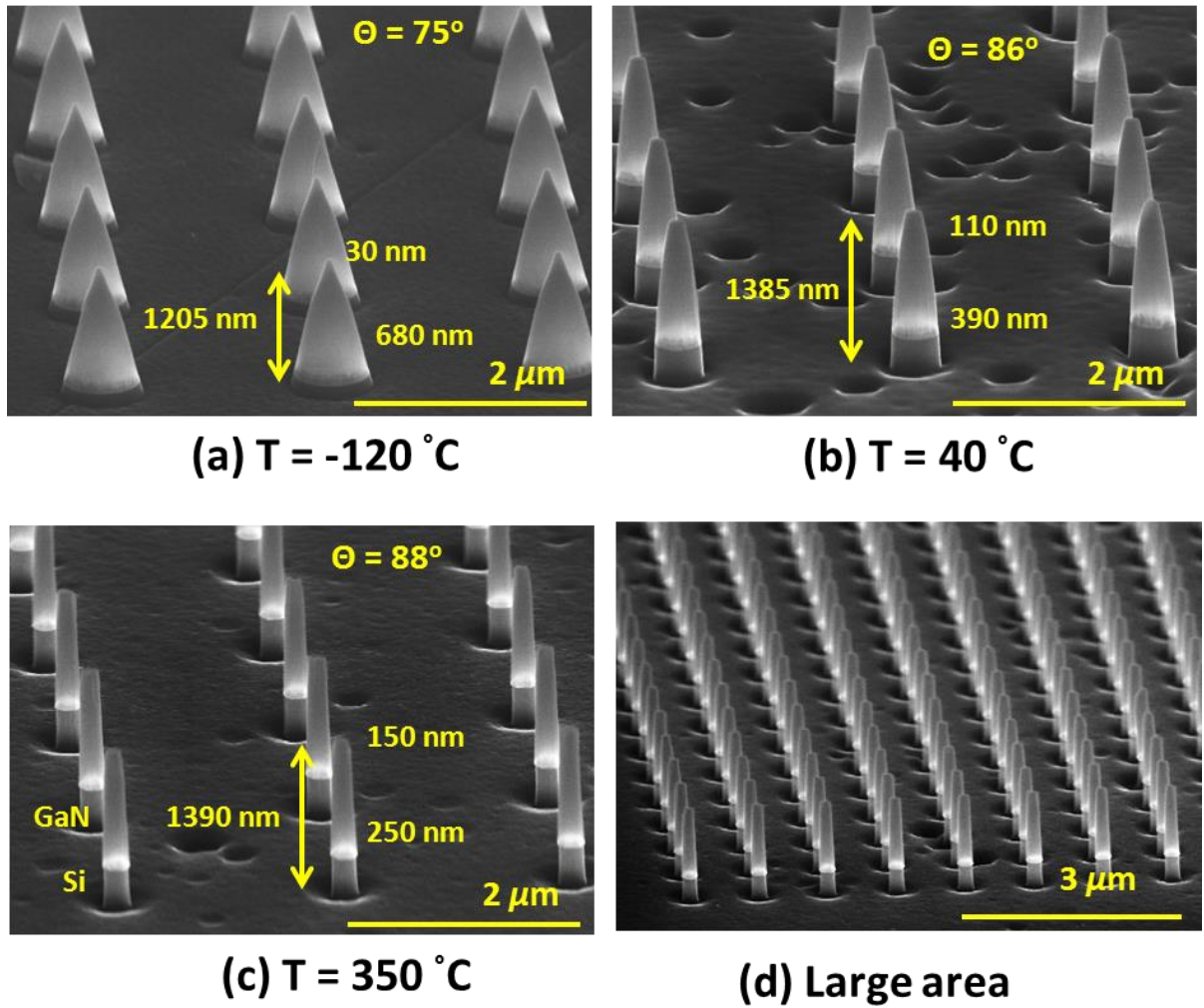

Fig.1. FESEM images of GaN nano pillars etched at diferrent substrate temperatures (a) -120 °C, (b) 40 °C and (c) 350 °C using $Cl_2/N_2/Ar$ (25/5/2 SCCM) chemistry with ICP power 800 W, RF power 300 W, pressure 0.66 Pa (5 mT) and etching time 5 min. The large uniformity of the nanopillars are shown in (d) for etching at 350 °C. The starting mask pattern was 250 nm diameter circle. Θ is labeled as the sidewall angle of the nanopillars. The height, top and bottom diameter of the nanopillars are also labeled inside the images.

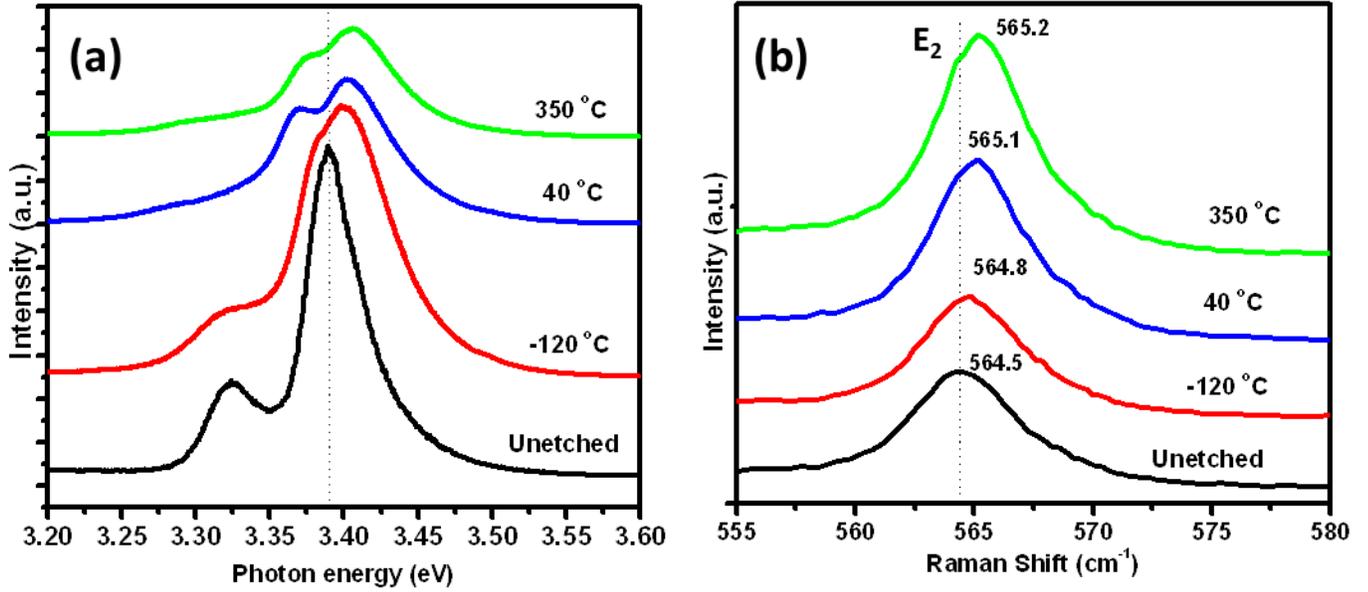

Fig.2. (a)Room temperature photoluminescence spectra and (b) room temperature Raman spectra for GaN nanopillars etched at different substrate temperature.

Fig. 2a. shows the room temperature micro PL spectra recorded from the GaN nanopillars fabricated at different temperature. Because of the much lower PL probing depth (~100 nm) at 325 nm excitation, the signal is primarily from the surface of the pillar arrays. The near-band-edge PL peak from the nanopillars is centered around 3.40 eV. The PL peak for the as grown and un-etched sample is centered at 3.39 eV. The band edge peak from the nanopillars shows a clear blue shift of about 1meV when compared with as grown GaN. A slight shift of PL peak position from 3.400 at -120 °C , 3.402 at 40 °C to 3.405 at 350 °C is observed from the nanopillars. This means pillars fabricated at higher temperature is more strain relaxed than the pillars fabricated at lower temperature. This blue shift is due to the relaxation from the tensile strain present in the GaN film on Si. Similar type of relaxation from compressive stress were observed by *Wang et al.*[17] and *Demangeot el al.*[12] Their GaN film was grown on sapphire substrate and they observed redshift in the PL peak from the nanopillars fabcriated using RIE etching. The GaN film grown on sapphire substrate feels compressive stress, while GaN film grown on Si substrate feels tensile stress. The energy bandgap of a semiconductor is affected by the residual stress in film. A tensile stress will result in a decrease of energy band gap while a compressive strain causes an increase of the band gap. Using the value of proportionality factor K=21.2 ± 3.2 meV/ GPa[21] for the stress induced PL peak shift, a tensile stress relaxation of about 0.5±0.1 GPa is estimated. *Wang et al.*[17] also observed the same amount of compressive stress relaxation on the GaN pillars fabricated on sapphire(0001) substrate. The secondary peak in between 3.33 – 3.75 eV range is associated with the donor acceptor pair (DAP) transition, and has been observed in PL spectra of GaN grown on Si.[15,19-20] The increase in intensity and red shift of this secondary peak may be associated to the defect formation, N vacancies or Ga interstitials in the surface layer of the GaN pillars etched at different temperatures.[12,21] Since atomic mass of N (14.006 amu) is less compare to the atomic mass of Ga (69.723 amu), there will be preferential sputtering of N due to ion etching,[15,22] which can form N vacancies and defects. N vacancies on the GaN surface behave as native donors and could indeed contribute to the stronger donor peak observed in the near band edge emission as reported by *Perlin et al.*[23] Thus we can say that nanopillars fabricated at lower temperature at -120 °C had lesser N vacancies and were more defect free compare to the pillars fabricated at higher

temperatures at 40 °C and 350 °C. Room Raman spectra from the nanopillars fabricated at different temperature are shown in fig. 2b. A systematic shift of the $E_2$ mode towards higher wavenumber with increasing temperature is observed here. This confirms that the nano pillar fabricated at higher temperature is more strained free in compare to that fabricated at low temperatures.

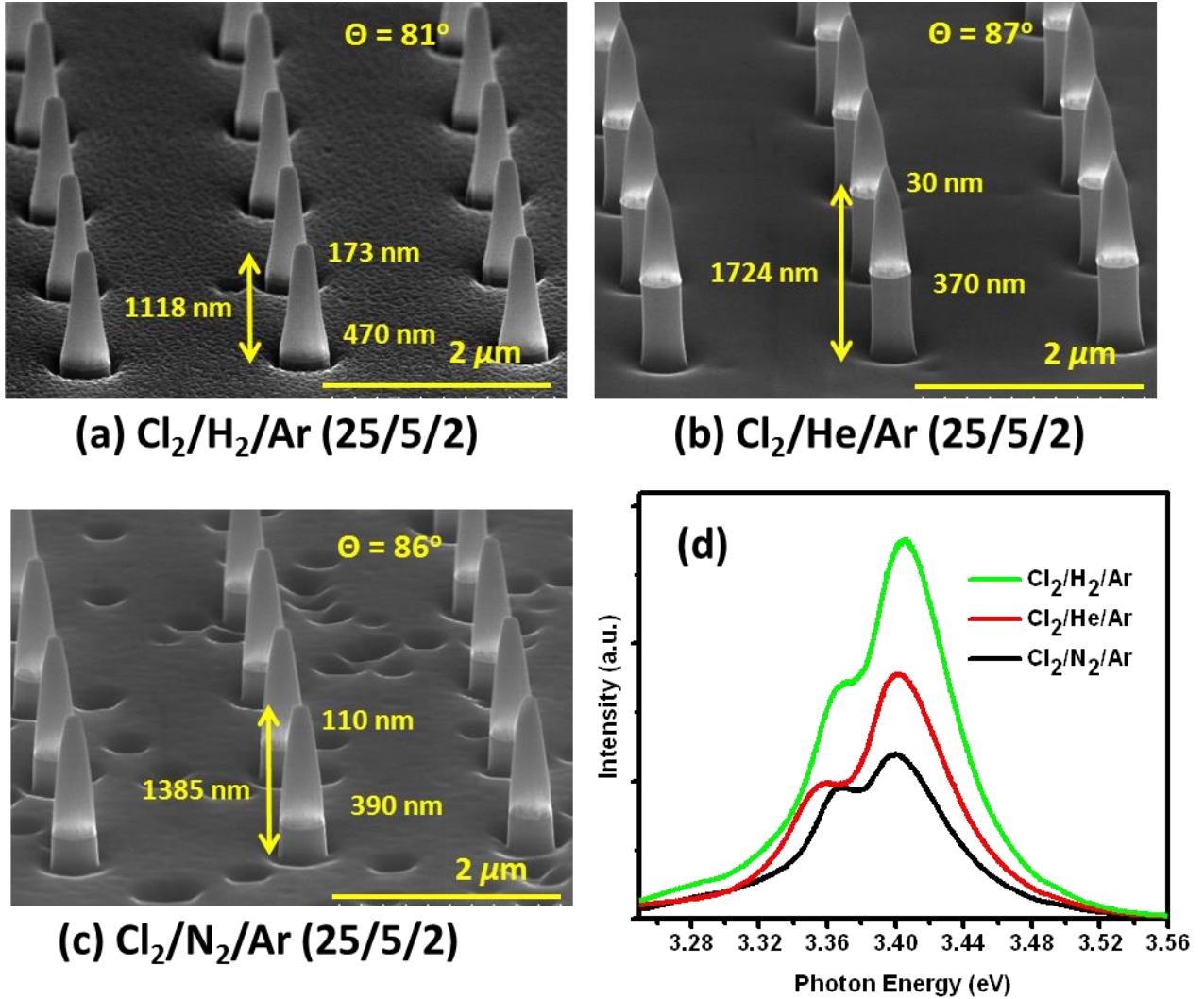

Fig.3. FESEM images of GaN nano pillars etched at different plasma chemistry (a) $Cl_2/H_2/Ar$ (25/5/2 sccm), (b) $Cl_2/He/Ar$ (25/5/2 sccm) and (c) $Cl_2/N_2/Ar$ (25/5/2 sccm) at 40 °C with ICP power 800 W, RF power 300 W, pressure 0.66 Pa (5 mT) and etching time 5 min. The starting mask pattern was 250 nm diameter circle. The room temperature photoluminescence spectra are shown in (d).

Figure 3 (a to c) shows the morphology of the nanopillars fabricated using different gas chemistries $Cl_2/H_2/Ar$ (25/5/2 sccm), $Cl_2/He/Ar$ (25/5/2 sccm) and $Cl_2/N_2/Ar$ (25/5/2 sccm) while keeping other parameters ICP power 800 W, RF power 300 W, temperature 40 °C, chamber pressure 0.66 Pa (5 mTorr) and time 5 min constant. Etch rate , and figure of merit for the nano pillar fabricated using different gas chemistry was reported in our earlier studies.[10] By measuring the height of the nanopollars at different etch conditions, it is clear that the etch rate is much less for $Cl_2/H_2/Ar$ (25/5/2 sccm) chemistry compare to that of $Cl_2/He/Ar$ (25/5/2 sccm), and $Cl_2/N_2/Ar$ (25/5/2 sccm). Suppression of GaN etch rate with addition of $H_2$ has been observed for $Cl_2/Ar$ plasma by Shul *et al*.[24] This was attributed to the consumption of reactive Cl radical by H forming HCl. Fig. 3d. shows the room temperature PL spectra from these nanopillars fabricated using different gas chemistries. The PL spectra shows no significant shift in the peak position. Nanopillars fabricated using $Cl_2/H_2/Ar$ (25/5/2 sccm) chemistry shows most intense PL peak compare to that etched using $Cl_2/He/Ar$ (25/5/2 sccm). This may be due to the fact that $H_2$ is a lighter atom compare to He and causes less damage to the nanopillar sidewalls. Obviously, the nanopillars fabricated using $Cl_2/N_2/Ar$ (25/5/2 sccm) would have much more surface damage compare to that in the case of previous two gas chemistries.

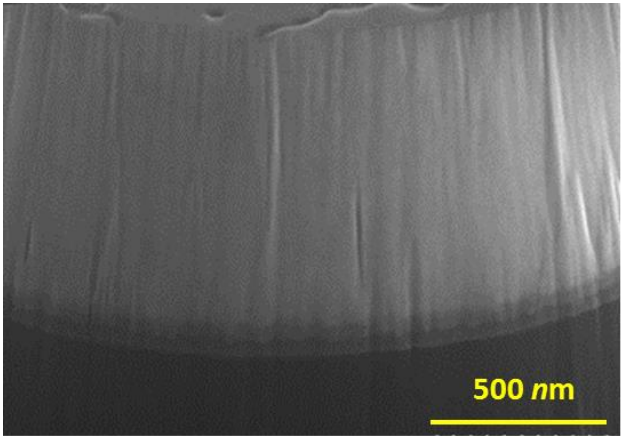
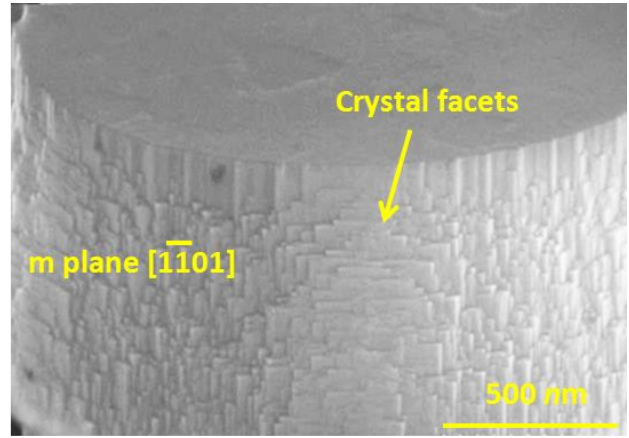
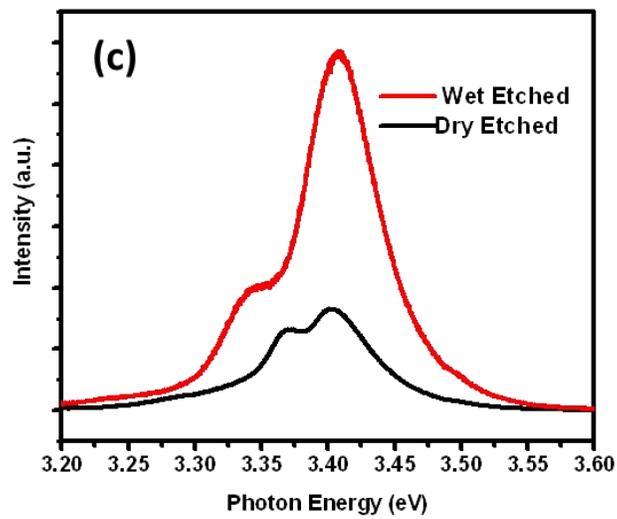

Fig.4. FESEM images of sidewalls of GaN nano pillars etched at after (a) ICP dry plasma etching using Cl2/N2/Ar (25/5/2 sccm) at 40 °C with ICP power 800 W, RF power 300 W, pressure 0.66 Pa (5 mT) and (b) after post etching wet chemical treatment to remove the sidewall damage using KOH solution. The (c) room temperature photoluminescence spectra shows the enhancement of the intensity from KOH wet etched sample compare to that from the ICP dry etched sample .

Fig.4 shows the morphology of the sidewall surface of GaN nanopillars after ICP dry etching and after wet chemical etching using 20 % KOH for 10 min at 40 °C. The KOH wet etching removes the damaged byproduct materials on the sidewall due to dry ICP etching. KOH etching forms crystalline facets on the side walls. This side wall plans are generally marked as non-polar m-plane [1$\bar{1}$01] for GaN. The advantage of nonpolar m plane GaN over polar c-plane[0001] GaN is that it does not have a large polarization charge. Large polarization charge in quantum well structures separates the electron and hole wave functions and thus decreases their recombination rate and overall efficiency.[25-26] Therefore it is technologically important to grow m-plane nonpolar GaN nano pillar structures. The KOH etching process normally attack the defective regions at the surface. KOH etching of GaN is polarity selective. The different etching characteristics of Ga-polar and N-polar face are due to the different states of the surface bonding. The N-face plane is well known to be more reactive and readily etched compare to the Ga-face. Exposure of the non-polar GaN surface to KOH first etches the N-face plane and then Ga-face is exposed by the recession of its adjacent N-face plan.[27-29] Thus the KOH wet etching produces surface texturing on the side wall of the GaN nanopillars as shown in fig. 4(b). An approximate 4 times increase in PL intensity is seen in Fig4.(c). The shape of the ICP dry etched and KOH wet etched PL spectra are similar, which indicate this texturing was able to extract light trapped in the crystal. This enhancement over a range of (3.25-3.55) of photon energies can be useful for application in surface photonic structures. The wet KOH treated surface in fig.4(c) show that the etch rate was uniform across the pitted and initially smooth region across the surface. The depth of the pits approximately 50 nm which corresponds to the depth of the quantum well region. Thus the textured surface on the side wall of the GaN nanopillar structures increases the effective surface area for the escape of the photons and enhanced the PL intensity from non-polar m-plane GaN LED structures.

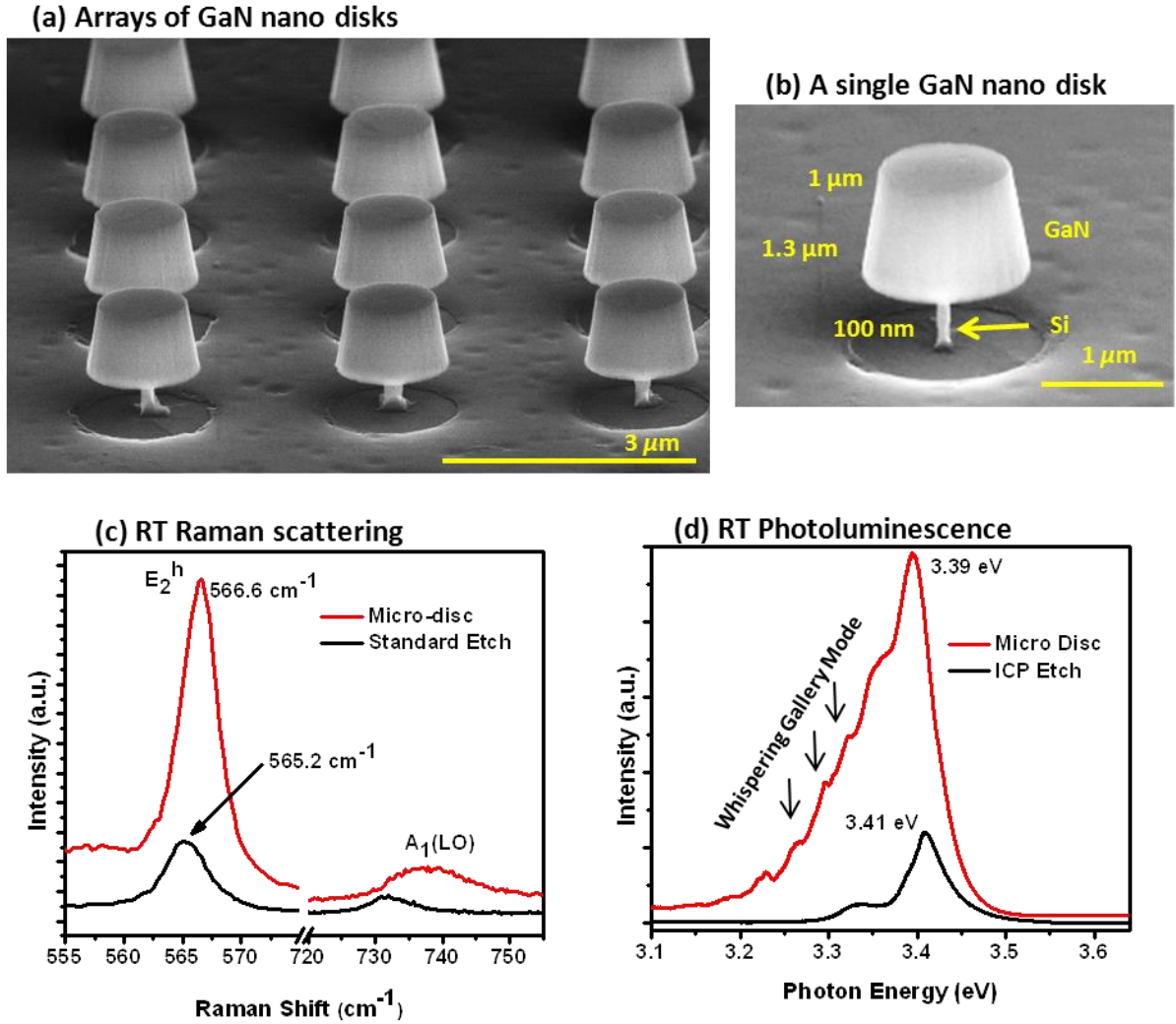

Fig.5. FESEM image of (a) arrays of GaN nano disks structure fabricated using ICP dry etch and post wet chemical etching by 10% KOH: EG solution, (b) a magnified single nano disk structure with top diameter 1μm, bottom diameter 1.3μ m and base silicon undercut diameter 100 nm. Room temperature (c) Raman scattering and (d) photoluminescence spectra from the dry etched nano pillars as well as from the nanodisk structure are also shown here.

Fig. 5a shows the arrays of free standing GaN nanodisk fabricated from the 1μm top diameter GaN/Si nanopillars after undercutting the Si with the help of wet chemical etching. The wet chemical etching was done using 10% KOH solution in ethylene glycol (EG) at 40°C for 10 min. Fig 5b shows a single GaN nano disks structure, with top diameter 1μm and bottom diameter 1.3μm, standing on top of Si undercut base pillar of 100 nm diameter. From the above figures it is clear that we have fabricated GaN nano disks with small diameter, smooth sidewalls, good circularity and uniformity across the sample over a large area. *Tamboli et al.*[30-31] and *Heberer et al.*[32] reported fabrication of undercut microdisks using band gap selective photoelectrochemical (PEC) etching on sapphire substrate. *Vickness et al.*[33] reported the fabrication of Si undercut beneath 200μm GaN microdisks using XeF$_2$. *Choi et al.*[34] had used HNO$_3$, H$_2$O and NH$_4$F solution to undercut Si surrounding the 20μm GaN microdisks. However most of all those reported microsdisk structures are large in diameter in the range of 5μm to 200μm and the side wall of the structures are rough. Here we have used EG instead of water as solvent for KOH so that we are able to etch Si very slowly.[35] Etching rate of Si was very sensitive to the temperature of the KOH:EG solution. At temperature 40 °C of the 10 % KOH in EG solution the etching rate of Si was found to be 70 nm/ min while the etching rate GaN is negligible. However etching rate increases abruptly with increase of temperature. This selective and slow etching process gives a smooth and circularly nanodisk structures over the narrow silicon post as shown in fig. 5a, b. The room temperature Raman scattering measurement from these nanodisks are shown in figure 5c. For comparison the Raman scattering data from dry etched nano pillars and before wet etching have also been show here. In case of GaN, usually the $E_2^h$ and A$_1$(LO) modes are observed in the z(xx)$\bar{z}$ geometry. The $E_2^h$ Raman peak is generally used to estimate the in-plane strain in the GaN.[36-37] As a reference, the $E_2^h$ phonon peak measured from bulk strain free GaN sample was found at 567.2 cm$^{-1}$ in our earlier studies.[38] The $E_2^h$ peak position for GaN nanopillar and nanodisk structure was found to be at 565.2 and 566.6 cm$^{-1}$ respectively. This indicate that the nanodisk structure are more strain free compare to the nanopillar structure. As compare to the Raman spectrum of the GaN nanopillar structure, a large increase in the Raman scattering efficiency is clearly observed from the nanodisk strucutres. This may be due to the removal of amorphous layer from the side wall of the GaN nanopillar due to KOH etching during the fabrication of the nanodisk structure. This enhancement in Raman intensity may also be due to the more efficient coupling and multiple scattering of light due to the presence of an air gap beneath the GaN surface that offer a large dielectric contrast as reported by *Choi et al.*[34] The optical properties of this nanodisk structures were probed by room temperature photoluminescence with the excitation of continuous wave (CW) and very low power (2 mW/cm$^2$) 325 nm He:Cd laser. The PL spectra from the nanopillars structures and the nanodisk structures are shown in Fig. 5d. Compare to the nanopillar structures, the PL spectra from the nanodisks structures shows enhancement in the luminescence intensity and emergence of periodic oscillation on the lower energy side of the PL spectra. The presence of these weak periodic oscillations are the indication of whispering gallery mode (WGM) generated from the nanodisk structure. However to clearly view this WGM modes the nanodisks structures need to be excited with high power (>300W/cm$^2$) laser.[30] *Choi et el.*[34] observed the presence WGM mode in 20 μm GaN microdisks structures in low (4.3 K) temperature PL spectra using high power pulse laser excitation. The enhancement of these WGM modes at low temperatures by exciting with high power pulsed laser will be presented elsewhere. We assume that our small diameter nanodisks structures can be useful for GaN based resonant cavity devices and towards the application of low power nanolaser.

## IV. SUMMARY AND CONCLUSIONS

We have studied in detail the surface morphology and optical properties of arrays of GaN nanopillar structures fabricated using different dry etching conditions as well as after post wet chemical etching. These studies give a detailed insight into the consequences of damage induced due to dry ICP etching as well as healing of those damages by post wet chemical etching treatment. The objective of this paper was to establish ICP etching combined with post wet etching as a top-down tool for fabrication of high quality and high aspect ratio GaN nanopillar structures on Si substrate. We plan to utilize these structures for device fabrication.

## ACKNOWLEDGMENTS

The nanostructures were fabricated at the Nanofab clean room of the NIST Center for Nanoscale Science and Technology. The University of Maryland portion of the work was partially supported by the Defense Threat Reduction Agency, Basic Research Award # HDTRA1-10-1-0107.